\documentclass[aps,prl,floatfix,twocolumn,showkeys]{revtex4}
\usepackage{amsfonts}
\usepackage{amsmath}
\usepackage{amssymb}
\usepackage{graphicx}

\begin{document}

\title{Magnetoresistive memory in phase separated La$_{0.5}$Ca$_{0.5}$MnO$_{3}$}

\author{J. Sacanell$^{a,\footnote{Corresponding Author: J. Sacanell; Av. Gral Paz
1499, San Martin (1650), Pcia. de Buenos Aires, Argentina.\\ FAX:
(5411)6772-7121\\ sacanell@cnea.gov.ar.}}$, F. Parisi$^{a}$, P.
Levy$^{a}$, L. Ghivelder$^{b}$\\}

\address{$^{a}$ Departamento de F\'{\i}sica, CAC, CNEA, Buenos
Aires, Argentina.\\$^{b}$ Instituto de F\'{\i}sica, UFRJ, Rio de
Janeiro, Brazil.}

\begin{abstract}
We have studied a non volatile memory effect in the mixed valent
compound La$_{0.5}$Ca$_{0.5}$MnO$_{3}$ induced by magnetic field
(H). In a previous work [R.S. Freitas et al., Phys. Rev. B 65
(2002) 104403], it has been shown that the response of this system
upon application of H strongly depends on the temperature range,
related to three well differentiated regimes of phase separation
occurring below 220 K. In this work we compare memory capabilities
of the compound, determined following two different experimental
procedures for applying H, namely zero field cooling and field
cooling the sample. These results are analyzed and discussed
within the scenario of phase separation.
\end{abstract}


\maketitle

\newpage

Rare earth based manganese oxides, also known as manganites, have
been the focus of extensive research since the discovery of the
colossal magnetoresistance effect (CMR) \cite{jin}. The close
interplay between ferromagnetic double exchange \cite{zener} and
antiferromagnetic superexchange gives rise to extraordinary
properties. The most intriguing one is the existence of a phase
separated state, i.e. the coexistence of ferromagnetic metallic
(FM) and charge ordered (CO) insulating phases \cite{dagotto}.

The possibility to manipulate the relative fraction of the
coexisting phases has been widely studied using several
techniques. Phase separation can be altered by introducing
chemical disorder \cite{uehara}, by changing the ceramic grain
size of the samples \cite{levyPRBlaca,Podzorov}, by applying
external hydrostatic pressure \cite{presion}, by the application
of external electric \cite{tokura} and magnetic fields
\cite{parisi,correlation} and  by thermal cycles of the samples
passing through a first order phase transition \cite{LowTirrev}.

In previous works \cite{nonvolatile,persistent} we have explored
the possibility of controlling the relative amount of the
coexisting phases in La$_{5/8-x}$Pr$_{x}$Ca$_{3/8}$MnO$_{3}$
(x=0.3, LPCMO) and
La$_{0.5}$Ca$_{0.5}$Mn$_{1-y}$Fe$_{y}$O$_{3}$(LCMFO) by the
application of magnetic field.

La$_{0.5}$Ca$_{0.5}$MnO$_{3}$ (LCMO) is one of the most studied
system in the literature of manganite materials. The compound,
paramagnetic at room temperature, changes on cooling to a mainly
FM metallic phase at $T_{C}\approx$ 220 K, and subsequently to
charge-ordered antiferromagnetic (CO-AFM) phase at $T_{co}\approx$
150 K (180 K upon warming)\cite{schiffer}. However, it has been
established that this system is better described as magnetically
phase-segregated over a wide range of temperatures
\cite{levyPRBlaca,freitas}, a phenomenon called phase separation
(PS). At low temperatures, T $< T_{co}$, FM metallic regions are
trapped in a CO-AFM matrix, whereas at an intermediate temperature
range ($T_{co} < T < T_{C}$), the FM phase coexists with
insulating non-FM regions.

It was established that three ranges can be identified in which
this phase coexistence exhibits different features under the
application of H \cite{freitas}:

- A soft PS state for 200 $K < T <$ 220 K: FM clusters coexist
with paramagnetic regions. In this range, $\rho$ is reversible
against the application of moderate magnetic fields.

- An intermediate PS for 150 $K < T <$ 200 K: Coexistence of FM
and insulator regions. The FM phase is partially confined but can
grow against the insulating one while applying a low H.

- A hard PS state for T $<$ 150 K: Coexistence of FM and CO-AFM
regions. The FM phase is structurally confined and cannot grow
against CO in moderate H.\\

In this report we have studied the possibility of imprinting
different values of H by inducing the irreversible growth of FM
and CO regions by applying magnetic fields in the field cooling
(FC) and zero field cooling (ZFC) modes in different temperature
ranges. Their relative amount acts as a sort of analogical memory
of the previously applied H. To recover the H value, M or $\rho$
measurements can be used.\\

Polycrystalline samples of La$_{0.5}$Ca$_{0.5}$MnO$_{3}$ were
used, their preparation procedure and structural characterization
are described elsewhere \cite{levyPRBlaca}. Transport measurements
were performed with the standard four probe method. Magnetization
measurements were performed in a commercial magnetometer (Quantum
Design PPMS).\\

Measurements were made using the the ZFC and the FC procedures: \\

In the ZFC mode the sample is cooled to the desired T in a field
H$_0$ = 0 for $\rho$ measurements. For M measurements a ZFC-like
procedure was performed cooling in a small H$_0$ $\approx$ 0.1 T.
Then, different H$_{ap}$ ($> H_0$) are turned on and off during
periods of tens of minutes, increasing H$_{ap}$ on each run. We
expect to write the different values of H$_{ap}$, in the relative
growth of the FM phase. In this case, H$_{ap}$ is the
perturbation.

Measurements in the ZFC mode were performed in the 150 K - 200 K
range, where we expect an irreversible change of the properties
\cite{freitas}.

Figure 1 (a) and (b) shows M and the applied H vs time and $\rho$
and the applied H vs. time respectively, for experiments performed
in the ZFC procedure at 170 K.

Sudden increases of M and decreases of $\rho$ are observed when
H$_{ap}$ is applied. These jumps are similar to the ones observed
in LPCMO and LCMFO \cite{nonvolatile,persistent} and are related
to the (fast) alignment of spins and domains and to the (slow)
enlargement of the FM phase. Remarkably, once H$_{ap}$ is turned
off, M decreases and $\rho$ increases without recovering their
previous H$_0$ values. Thus, the presence of a persistent effect
directly related to the magnitude of the previously H$_{ap}$ is
apparent.

This effect was found in the 150 K - 200 K temperature range,
while below T$_{co}$ it is almost negligible.

In agreement with previous results on LPCMO and LCMFO
\cite{nonvolatile,persistent}, we can relate this effect to the
increase of the FM fraction ($f$) of the sample every time an
increasing H pulse is applied. However, we have to note that the
effect is less significant in the present work than the one
observed in the above mentioned references.

By using M measurements, the relative change of $f$ after the
whole experiment and the application of H = 6 T can be estimated
by the ratio of the "initial" and "final" values of magnetization
under no perturbation, indicated in the figure as M$_i$ and M$_F$.
With these values, we obtain $f_F/f_i \approx$ 1.10, i.e. an
increase of 10 \% in the FM fraction.

Assuming a linear dependence of $\Delta f$ with H, this would mean
an increase of less than 2 \% of $f$ by Tesla. Also, from fig.
1(b), we can estimate MR / H $\approx$ 5\%/Tesla, a very low value
when compared to the almost 80\%/Tesla achieved by LPCMO
\cite{nonvolatile}.\\

As LCMO exhibits robust CO features in a broad T range, we have
explored an alternative way to produce memory effects, inducing
the growth of CO regions by field cooling the sample (FC mode).\\

\begin{center}
\begin{figure}
  \includegraphics[height=11 cm]{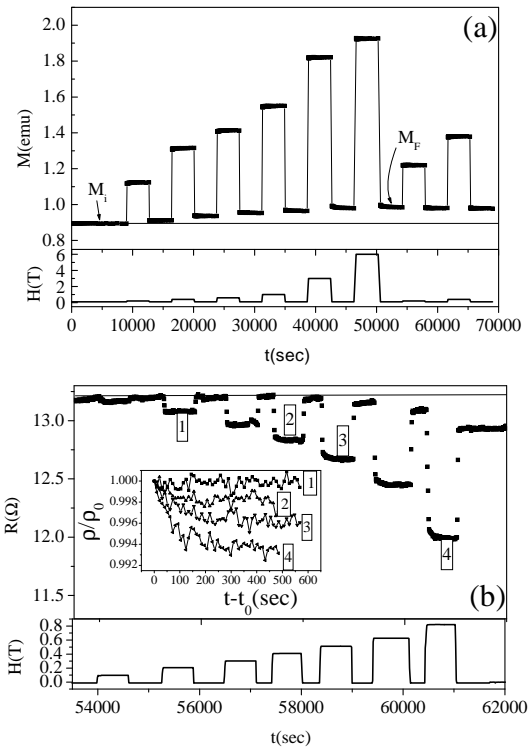}\\
  \caption{ZFC procedure: (a)M(170K) vs time for LCMO upon the application of H = 0.2, 0.4,
  0.6, 1, 3, 6, 0.2 and 0.4 T over H$_0$. (b) R(170 K) vs time for LCMO upon the
  application of H = 0.1, 0.2, 0.3, 0.4, 0.5, 0.6 and 0.7 T. Inset:
  normalized $\rho$(170 K) vs time during the onset of H. Number labels
  correspond to the ones in main figure.}\label{Fig1}
\end{figure}
\end{center}

In the FC mode the sample is cooled to the desired T in H$_{FC}
\approx$ 1 T. We have measured $\rho$ and M while applying short
"negative" field pulses, i.e. reducing the magnetic field to
H$_{ap}$ $< H_{FC}$ and returning after each pulse to $H_{FC}$.
The perturbation in this procedure is given by pulses of $\Delta$H
= H$_{FC}$ - H$_{ap}$. In this case, we expect the compound to
memorize the applied $\Delta$H value in the growth of non - FM
regions.

FC measurements were performed below T$_{co}$, where the CO is
harder.

Experiments performed at T = 130 K are shown in fig. 2 (a) and
(b), where we see that a much larger effect can be obtained by
making this simple modification of the ZFC experimental procedure.
In the ZFC mode a change of around 10 \% was observed in $f$ as
mentioned above, while for the FC mode, a change of almost 40 \%
can be achieved after the application of $\Delta$H $\approx$ 2 T
as can be obtained from fig. 2.

The application of the field pulses in the FC procedure to the
sample (see fig. 2), result in a huge reduction of the M value and
an equivalent increase of $\rho$, after returning to H$_{FC}$.

The change observed in this last experiment is related to the
increase of the CO phase during the temporary reduction of the
field.

The applied $H_{FC}$ forces a FM state in regions that otherwise
would be CO. The subsequent return to $H_{FC}$, yields an effect
similar to that obtained for the FM regions in the ZFC procedure.

The inset of fig. 2(b) shows the relaxation of the system after
the field returns to H$_{FC}$. This relaxation cannot be seen on M
measurements due to the relatively small magnitude of the changes
obtained, but is clear on resistivity ones. After reducing the
field in $\Delta$H = 0.1 T (labeled as relax. 1), relaxation is
towards an increment in $\rho$. After $\Delta$H = 0.2 T (labeled
as relax. 2), the system do not show relaxation and on subsequent
increases of $\Delta$H, $\rho$ reduces with time. This last result
indicates the critical value $\Delta$H$_{crit}$ above which the CO
phase attains an overenlarged metastable state, and signs the
existence of an equilibrium value of the FM (or CO) fraction which
depends on H$_{FC}$.\\

\begin{center}
\begin{figure}
  \includegraphics[height=11 cm]{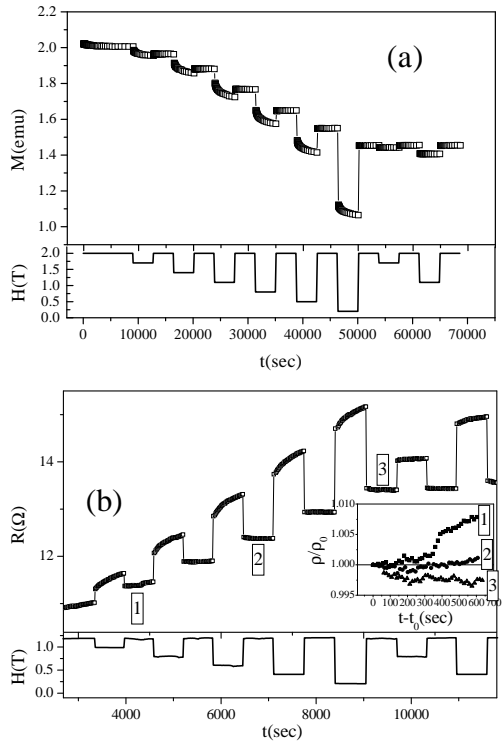}\\
  \caption{FC procedure: (a)M(130K) vs time for LCMO upon the reduction of H in
  0.3, 0.6, 0.9, 1.2, 1.5, 1.8, 0.3 and 0.9 T (H$_{FC}$ = 2 T).
  (b) R(130 K) vs time for LCMO upon the reduction of H in 0.1, 0.2, 0.3, 0.4,
  0.5, 0.2 and 0.4 T (H$_{FC}$ = 0.7 T).
  Inset: normalized $\rho$(130 K) vs time after H has returned to H$_{FC}$.
  Number labels correspond to the ones in main figure.}\label{Fig2}
\end{figure}
\end{center}

As a common feature for both procedures, we can observe that the
memory effect related to changes in the relative fraction of the
phases, is persistent after the "perturbation" (increase in H in
the ZFC mode, decrease in H in the FC mode) is turned off, and can
only be modified if an ultherior higher perturbation is applied.

In both procedures we can see that alignment and enlargement
effects are present. We argue that the enlargement of the FM phase
in the ZFC mode and of the CO phase in the FC mode is the
responsible of the memory effect. Because of this, if we apply a
smaller perturbation than the last applied, only alignment effects
are expected, as actually happens (see fig. 1 and 2).

Additionally, we can make an estimation of the change in $f$ in
the FC procedure using M and $\rho$ measurements.

For M measurements, the same procedure as presented for the ZFC
mode can be used. For resistivity measurements, we can make an
indirect determination of $f$ using a semi phenomenological model
for transport through a binary mixture known as General Effective
Medium theory or GEM \cite{gem}, which provides a relation to
obtain the samples' resistivity ($\rho_e$) as a function of the
resistivities of the constitutive phases (see refs.
\cite{parisi,correlation}).

From our M measurements, we obtain $\Delta f$/H $\approx$
16\%/Tesla, while for $\rho$, the value slightly changes to
14\%/Tesla.\\

Summarizing, we have observed a persistent effect that can be used
as an analogical memory of previously applied magnetic fields. The
effect is related to the enlargement of the FM or the CO phase
according to the particular procedure (ZFC or FC). We have shown
that temperature is very significant and has to be taken into
account when studying this effect. A previous characterization of
the system (as in \cite{freitas} for LCMO) is useful to choose the
appropriate temperature range to obtain a more sensitive response
of the system. We have found that, for LCMO, FC seems to be the
better procedure because larger changes can be observed in both M
and $\rho$ for changes in H of the same order of magnitude. We
have also observed that this effect is enhanced in the range in
which the CO phase is harder (see \cite{freitas}).\\


\begin{thebibliography}{99}
\bibitem[1]{jin} S. Jin et al., Science 264, (1994) 413.

\bibitem[2]{zener} C. Zener, Phys. Rev., 81, (1951) 440;
C. Zener, Phys. Rev., 82, (1951) 403.

\bibitem[3]{dagotto} E. Dagotto et al.,Phys. Rep. 344 (1-3) (2001)
1.

\bibitem[4]{uehara} M. Uehara et al, Nature 399 (1999) 560.

\bibitem[5]{levyPRBlaca} P. Levy et al., Phys. Rev.  62, (2000) 6437.

\bibitem[6]{Podzorov} V. Podzorov et al, Phys. Rev. B 64 (2001)
140406(R).

\bibitem[7]{presion} K. Khazeni et al., Phys. Rev. Lett. 76 (2), (1996) 295;
V. Laukhin et al., Phy. Rev. B 56 (16), (1997) 10009(R).

\bibitem[8]{tokura} A. Asamitsu et al., Nature 388, (1997) 50.

\bibitem[9]{parisi} F. Parisi et al., Phys.Rev. B 63 (2001) 144419.

\bibitem[10]{correlation} J. Sacanell et al, Phys. B: Cond. Matt. 320 (1-4)
(2002) 90.

\bibitem[11]{LowTirrev} J. Sacanell et al., Jour. of Alloys and Compounds 369
(2004) 74-77.

\bibitem[12]{nonvolatile} P. Levy et al., Phys. Rev. B 65 (2002) 140401(R).

\bibitem[13]{persistent} P. Levy et al., Jour. of Mag. and Mag.
Mat. 258-259 (2003) 293.

\bibitem[14]{schiffer} P. Schiffer, et al., Phys. Rev. Lett. 75 (1995)
3336.

\bibitem[15]{freitas} R.S. Freitas et al, Phys. Rev. B 65 (2002) 104403.

\bibitem[16]{gem} D.S. McLachlan, J. Phys. C: Solid State Phys. 20, (1987) 865.

\end{thebibliography}
\end{document}